\documentclass[10pt,elsart,elsart12,epsfig,epsf,graphics,longtable,amstex,amssymb,euscript]{article}
\usepackage{epsfig}
\usepackage{amssymb}

\begin{document}

\date{}
\title{Backgrounds in Neutrino Appearance Signal at MiniBooNE}

\author{Zelimir Djurcic\footnote{zdjurcic@nevis.columbia.edu} \hspace{2mm} for the MiniBooNE collaboration \footnote{This work is supported by NSF and DOE. Collaboration list is available at www-boone.fnal.gov.} }

\maketitle

\begin{abstract}
The MiniBooNE (Booster Neutrino Experiment) experiment is a precise
search for neutrino oscillations but is crucially sensitive to
background processes that would mimic an oscillation signal.
The background sources include intrinsic $\nu_{e}$ from kaon and
muon decays, mis-identified $\pi^{0}$, gammas from radiative 
delta decays, and muon neutrino events mis-identified as electrons. 
A summary of these backgrounds and the tools to handle them is presented.
\end{abstract}

\section{Introduction}
The MiniBooNE experiment will confirm or refute the LSND result \cite{lsnd}
with higher statistics and different sources of systematic error.
While LSND observed an excess of $\bar\nu_{e}$ events in a $\bar\nu_{\mu}$ beam,
the MiniBooNE is a $\nu_{\mu} \rightarrow \nu_{e}$ search.
The MiniBooNE detector is located 541 $m$ from a neutrino source
on the Booster neutrino beam line at Fermi National Accelerator
Laboratory. Details of the beam line, the detector, major 
physics interactions and a blind-analysis strategy can be found 
in Ref.\cite{conrad_and_brice}.
The interaction point, event time, energies, and the 
particle tracks are recorded from the times and charges of the PMT hits
in the detector. The detector is calibrated over the full energy range from 50
to 1000 $MeV$ using all event types. Stopping cosmic ray muons are used to calibrate
the energy scale for muon-type events and measure the position
and angle reconstruction resolution, when their path length can
be identified. This is accomplished with a scintillator hodoscope
on the top of the detector, combined with scintillation cubes at
various positions within the detector volume. Observed Michel
electrons from muon decay are used to calibrate the energy scale
and reconstruction resolution of electron-type events at the
52.8 $MeV$ Michel endpoint. Reconstructed $\pi^{0}$ events 
provide another electron-like calibration source. The 
photons that are emitted in $\pi^{0}$ decays span a
considerable range to over 1000 $MeV$. The $\pi^{0}$ mass
derived from reconstructed energies and directions
of two $\gamma$-rays has a peak at 136.3$\pm$0.8 $MeV$.
This is in an excellent agreement with 135.0 $MeV$ expectation,
providing a check on the energy scale and the reconstruction
over the full energy range of interest for $\nu_{e}$
appearance analysis.  

The events described in the calibration procedure are used as a basis
for a particle identification (PID). 
PID is performed by different algorithms that use the difference 
in characteristics of the Cerenkov rings and scintillation light 
associated with electrons, muons, protons or $\pi^{0}$'s.
These algorithms include a maximum likelihood method, neural net,
and boosted decision trees \cite{pid_paper}. 
The left part of Fig.~\ref{fig:fig2} shows an example of a boosted decision
tree where the muon/electron separation was measured with cosmic ray
muons and associated electrons. The overlap region in the electron distribution 
defines a fraction of muons mis-identified as electrons. Similar PID 
separation is performed between electrons and other types 
of the detected events.

\section{The Backgrounds in the Appearance Signal}

The beam that arrives at the detector is almost pure $\nu_{\mu}$ with a 
small (0.6\%) contamination of $\nu_{e}$ coming from muon and kaon decays 
in the decay pipe. 
The $\nu_{e}$ from $\mu$-decay
are directly tied to the observed $\nu_{\mu}$ interactions. Taking
into account a small solid angle subtended by the MiniBooNE detector,
the pion energy distribution can be determined from the energy of
the observed $\nu_{\mu}$ events. The pion energy spectrum is then
used to predict the $\nu_{e}$ from $\mu$-decay.
A second source of $\nu_{e}$ originates from $K_{e3}$-decay. This
component will be constrained using data from two
low energy production experiments: HARP experiment at CERN \cite{harp} and E910
experiment at BNL \cite{e910}. An additional check will be performed by a system 
called LMC (Little Muon Counters), measuring the muons from pion and kaon decays 
coming at wide angles from the decay pipe \cite{lmc}.
\begin{figure}[!h]
 \resizebox{1.0\columnwidth}{!}
 {\includegraphics[width=5.cm, height=3.cm]{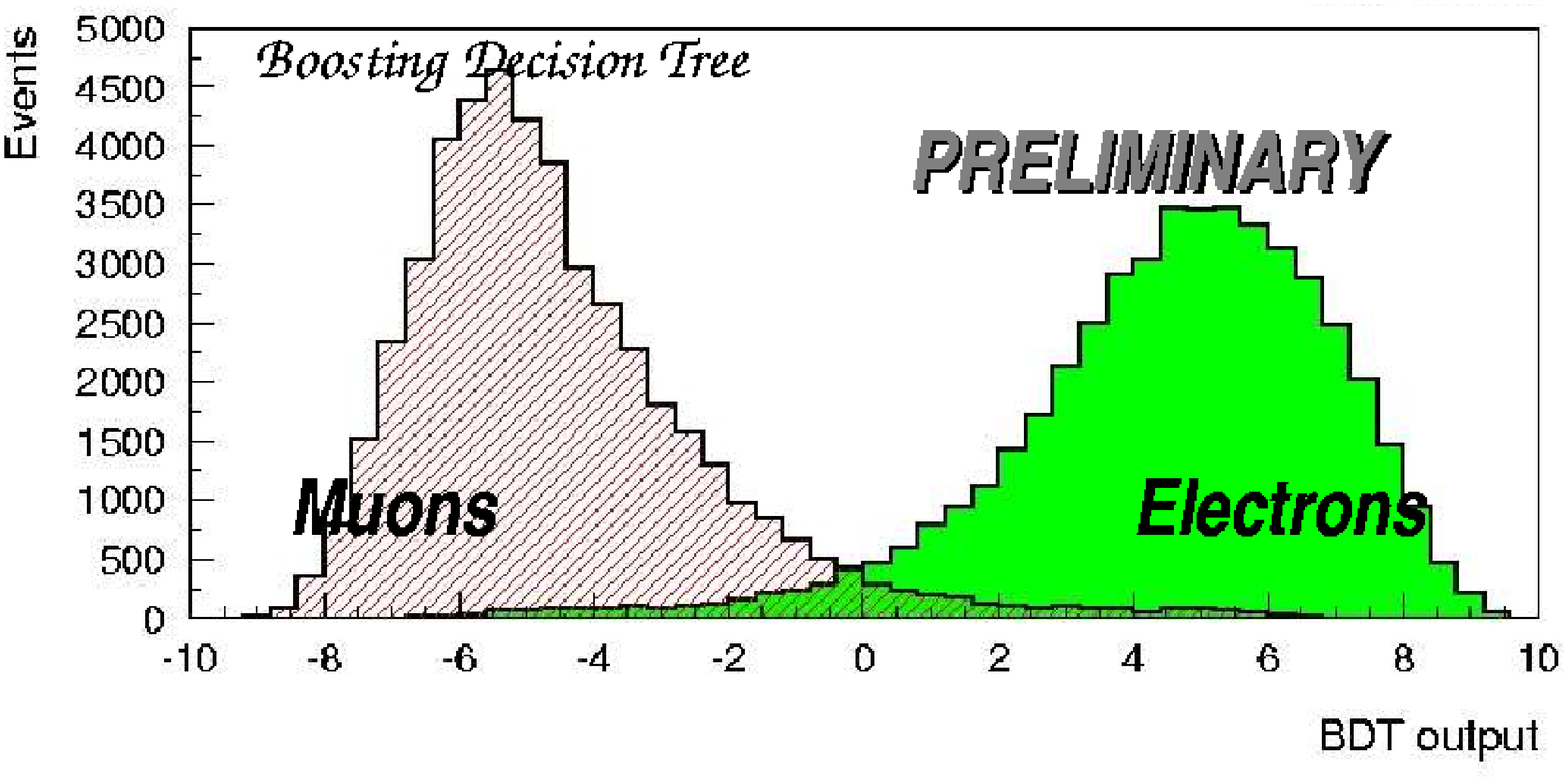}\includegraphics[width=6.cm, height=2.9cm]{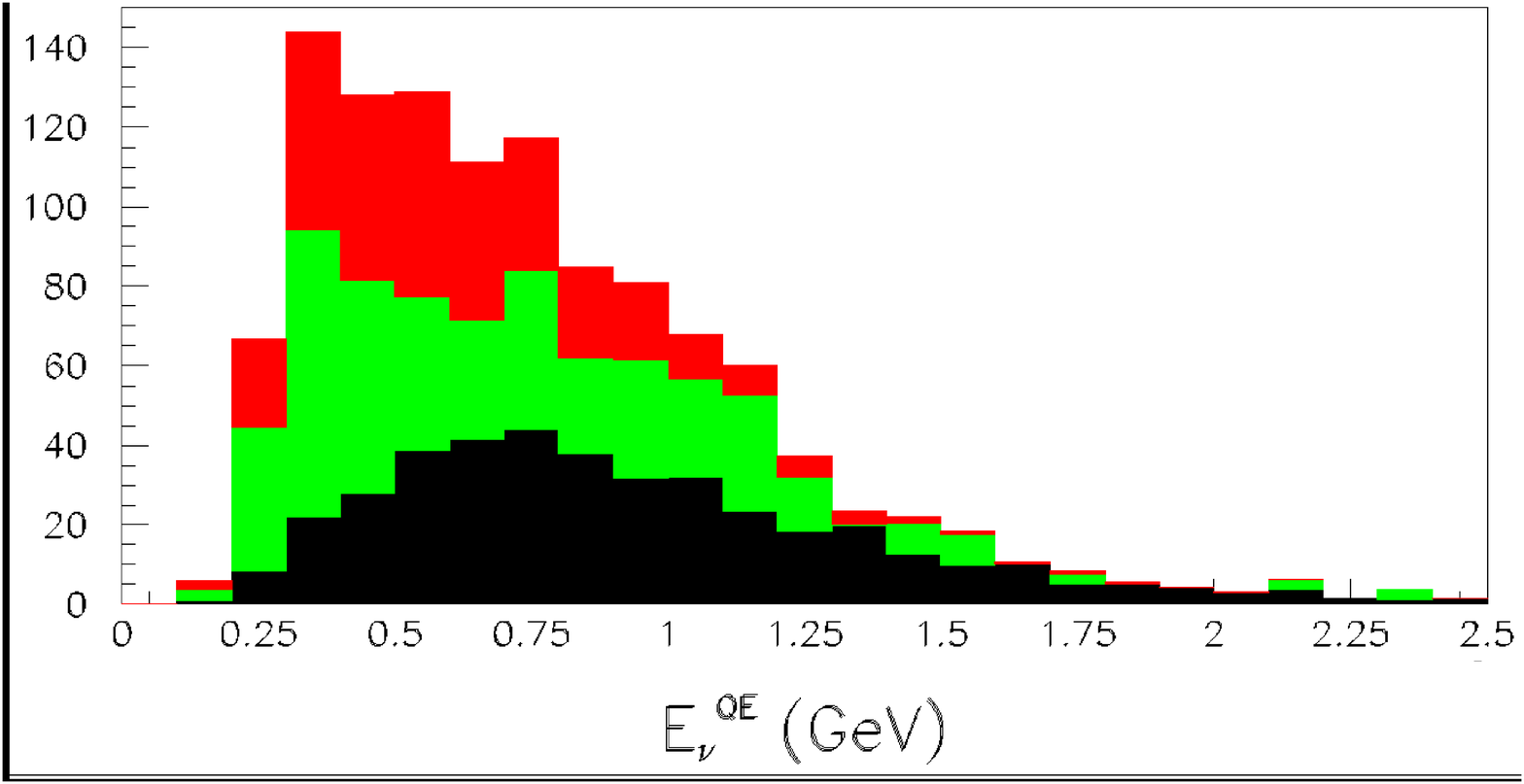}}
 \caption{Left: PID separation between stopped cosmic-ray muons (hatched region) and their decay (Michel) 
electrons (solid region) as obtained from boosted decision tress (BDT) using the MiniBooNE data. 
Right: Summed Monte Carlo event energy distribution for the oscillation signal 
($\Delta m^{2} = 0.4 eV^{2}$ and $\sin^{2}2\theta = 0.02$) (top), mis-identified $\pi^{0}$ events (middle)
and the intrinsic $\nu_{e}$ background events (bottom).}
 \label{fig:fig2}
\end{figure}
An important background component comes
from $\nu_{\mu}$ neutral production of $\pi^{0}$. Most $\pi^{0}$
are identified by the reconstruction of two Cherenkov rings produced
by two decay $\gamma$-rays. However, the decay of $\pi^{0}$ 
can appear much like primary electron emerging from a $\nu_{e}$
charged current interaction if one of the gammas from the decay 
overlaps the other, or is too low in energy to be detected.
Over 99\% of the NC $\pi^{0}$ are rejected in the appearance analysis.

\begin{figure}[!h]
 \resizebox{1.0\columnwidth}{!}
   {\includegraphics[width=12.cm, height=3.cm]{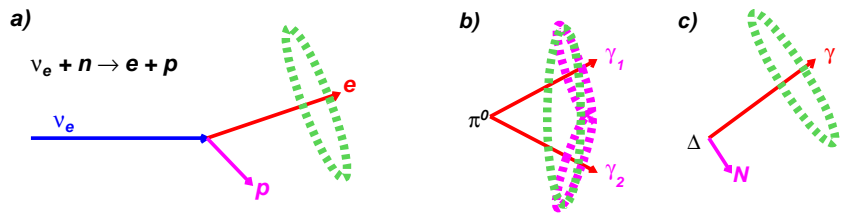}}
 \caption{An illustration showing: a) the signal, b) $\pi^{0}$ background, c) $\Delta$ background
in the events being detected through the Cherenkov ring reconstruction in the detector.}
 \label{fig:fig1}
\end{figure}

In addition to its primary
decay $\Delta \rightarrow \pi\;N$, the $\Delta$ resonance has a branching
fraction of 0.56\% to $\gamma \; N$ final state. The $\gamma$-ray may mimic an
electron from $\nu_{e}$ interaction. The rate of $\Delta$ production in neutral
current interactions can be estimated from the data, using the sample of
reconstructed $\pi^{0}$ decays. Fig.~\ref{fig:fig1} illustrates the imitation
of the signal by $\pi^{0}$ and $\Delta$ events.

Most $\nu_{\mu}$ events can be
easily identified by their penetration into the veto region when exiting muons
fire the veto, or by muons stopping in the inner detector and producing a Michel 
electron after a few microseconds.
\begin{table}
\begin{tabular}{ll}
\hline
\multicolumn{1}{l}{Event Class}  & \multicolumn{1}{l}{Cross-check} \\  \hline
$K^{+}$               & HARP, LMC, External Data\\ 
$K^{0}$               & E910, External Data\\
$\mu$                 & MiniBooNE Data\\
$\pi^{0}$             & NuMI, MiniBooNE Data\\
Other ($\Delta$, etc) & NuMI, MiniBooNE Data  \\ \hline
\end{tabular}
\caption{The MiniBooNE analysis is verified by different experimental cross-checks 
for each event class relevant to $\nu_{e}$ appearance search. }
\label{tab:crosscheck}
\end{table}

Another handle on the background comes from the energy and/or PID output 
distributions of the signal and various background components. Figure~\ref{fig:fig2} (right)
shows the energy distribution of the background summed with an oscillation 
signal for $\Delta m^{2} = 0.4 eV^{2}$ and $\sin^{2}2\theta = 0.02$.
A fit to the energy distribution results in an enhanced sensitivity as the intrinsic $\nu_{e}$ events
show a high energy tail while mis-identified $\pi^{0}$ are much narrower.

An important cross-check of electron event reconstruction and particle
identification comes from NuMI events observed in the MiniBooNE detector \cite{aaa}.
NuMI events consist of $\nu_{e}$, $\nu_{\mu}$, $\pi^{\pm}$, $\pi^{0}$ and $\Delta$
over the range of the energies relevant to the appearance analysis in MiniBooNE. 
A complete list of cross-checks is given in Table~\ref{tab:crosscheck}.

\end{document}